\documentclass[a4paper,11pt]{article}

\usepackage{amsfonts}
\usepackage{amsmath}
\usepackage{amssymb}
\usepackage{authblk}
\usepackage{cancel}
\usepackage[font={footnotesize,it}]{caption}
\usepackage{cite}
\usepackage{color}
\usepackage{comment}
\usepackage{default}
\usepackage{enumitem}
\usepackage{epsfig}
\usepackage{float}
\usepackage[margin=2.5cm]{geometry}
\usepackage{graphicx}
\usepackage{hyperref}
\usepackage{cleveref} 
\usepackage[utf8]{inputenc}
\usepackage{mathtools}
\usepackage{ragged2e}
\usepackage{subcaption}
\usepackage{tikz}

\hypersetup{
citecolor=red,
colorlinks=true,
filecolor=red,
linkcolor=blue,
linktocpage=true,
urlcolor=blue
}

{\makeatletter \g@addto@macro\bfseries{\boldmath} \makeatother}

\usetikzlibrary{shapes}

\begin{document}

\title{Covariant holographic entanglement negativity for disjoint intervals in $AdS_3/CFT_2$}

\author[1]{Vinay Malvimat\thanks{\noindent E-mail:~ {\tt vinaymmp@gmail.com}}}
\author[2]{Sayid Mondal\thanks{\noindent E-mail:~ {\tt sayidphy@iitk.ac.in}}}
\author[2]{Boudhayan Paul\thanks{\noindent E-mail:~ {\tt paul@iitk.ac.in}}}
\author[2]{Gautam Sengupta\thanks{\noindent E-mail:~ {\tt sengupta@iitk.ac.in}}}

\affil[1]{
Indian Institute of Science Education and Research\\

Homi Bhabha Rd, Pashan, Pune 411 008, India
\bigskip
}

\affil[2]{
Department of Physics\\

Indian Institute of Technology\\ 

Kanpur 208 016, India
}

\date{}

\maketitle

\thispagestyle{empty}

\begin{abstract}

\noindent

\justify

We advance a construction for the covariant holographic entanglement negativity for time dependent mixed states of disjoint intervals in $(1+1)$ dimensional conformal field theories ($CFT_{1+1}$) dual to bulk non static $AdS_3$ geometries. Application of our proposal to such mixed states in a $CFT_{1+1}$ dual to bulk non extremal and extremal rotating BTZ black holes exactly reproduces the replica technique results in the large central charge limit. We also investigate the time dependent holographic entanglement negativity for such mixed states in a $CFT_{1+1}$ dual to a bulk Vaidya-$AdS_3$ geometry in the context of their thermalization involving bulk black hole formation.

\end{abstract}

\clearpage

\tableofcontents

\clearpage

\section{Introduction}
\label{sec_intro}

Recently quantum entanglement has emerged as one of the most active research areas, garnering interest from a diverse range of fields extending from condensed matter physics to quantum gravity. For bipartite pure states the entanglement may be characterized by the entanglement entropy which is defined as the von Neumann entropy of the reduced density matrix for the corresponding subsystem. This is dif{}ficult to compute for extended quantum many body systems with infinite degrees of freedom although a formal definition may be characterized through a replica technique. For $(1+1)$ dimensional conformal field theories however the entanglement entropy may be explicitly computed through the replica technique as described in \cite{Calabrese:2004eu,Calabrese:2009qy}, as a correlation function of the relevant branch point twist fields \cite{Cardy:2007mb} (see \cite{1987NuPhB.282...13D} and \cite{1987CMaPh.112..567K} for details on twist fields).

It is well known from quantum information theory that the entanglement entropy fails to characterize mixed state entanglement as it involves correlations irrelevant to the entanglement of the mixed state under consideration. This crucial issue was addressed in a significant communication by Vidal and Werner through the introduction of a computable measure termed entanglement negativity (logarithmic negativity) which characterized the upper bound on the distillable entanglement for such mixed states \cite{Vidal:2002zz}.\footnote{Distillable entanglement describes the amount of entanglement that can be extracted from a mixed state using only LOCC.}
This measure was defined as the logarithm of the trace norm of the partially transposed reduced density matrix of a bipartite system with respect to one of the subsystems. In \cite{Plenio:2005cwa} it was observed that this measure although not convex, was an entanglement monotone under local operations and classical communication (LOCC).

A variant of the replica technique was later developed to compute the entanglement negativity for various mixed state configurations in $CFT_{1+1}$s \cite{Calabrese:2012ew,Calabrese:2012nk,Calabrese:2014yza}. Subsequently this technique was extended to compute the entanglement negativity following a global quench for time dependent mixed states of disjoint and adjacent intervals in a $CFT_{1+1}$ \cite{Coser:2014gsa}. The time evolution of the entanglement negativity for certain mixed states in a $CFT_{1+1}$ following a local quench was subsequently investigated in \cite{Hoogeveen:2014bqa,2014NJPh...16l3020E,Wen:2015qwa}. 

In a seminal communication Ryu and Takayanagi (RT) \cite{Ryu:2006bv,Ryu:2006ef} advanced a holographic conjecture for the entanglement entropy of a subsystem in a $CFT_d$ dual to a bulk $AdS_{d+1}$ geometry in the $AdS/CFT$ scenario. The entanglement entropy was found to be proportional to the area of a co dimension two bulk static minimal surface anchored on the subsystem. The RT conjecture inspired intense activity in this area leading to several interesting insights \cite{Nishioka:2009un,Takayanagi:2012kg,Nishioka:2018khk} (and references therein). However the RT conjecture described above was valid only for $CFT_d$s dual to static bulk $AdS_{d+1}$ configurations. For non static (stationary or time dependent) configurations it was required to formulate the precise characterization of minimal surfaces in the context of a Lorentzian signature space time.

In \cite{Hubeny:2007xt} Hubeny, Rangamani and Takayanagi (HRT) addressed the critical issue described above and proposed a covariant holographic conjecture for the entanglement entropy in a $CFT_d$ dual to a non static bulk $AdS_{d+1}$ geometry. Their proposal involved the covariant Bousso entropy bound \cite{Bousso:1999xy,Bousso:1999cb,Bousso:2002ju} which results from a light sheet construction. In this non static scenario the holographic entanglement entropy of a subsystem in a $CFT_d$ could be described in terms of the area of a co dimension two bulk extremal surface anchored on the subsystem.\footnote{In the context of the $AdS_3/CFT_2$ correspondence these are bulk space like geodesics anchored on the corresponding spatial interval describing the subsystem.}
The HRT conjecture was subsequently utilized to obtain the entanglement entropy for various stationary and time dependent configurations in holographic $CFT$s \cite{Balasubramanian:2011ur,Hartman:2013qma,Albash:2010mv,Caputa:2013eka,Mandal:2016cdw,David:2016pzn}.

The developments described above leads to the significant issue of a holographic characterization for the entanglement negativity of mixed state configurations in a dual $CFT_d$. In \cite{Chaturvedi:2016rcn,Chaturvedi:2016rft} two of the present authors (VM and GS) advanced a holographic entanglement negativity conjecture for mixed state configuration of a single interval in $CFT$s through the $AdS_{d+1}/CFT_d$ correspondence and its covariant extension in \cite{Chaturvedi:2016opa}.
The above analysis led to interesting insights and reproduced universal features of entanglement negativity from a holographic perspective. The conjecture was substantiated through a robust consistency check for the $AdS_3/CFT_2$ scenario involving a large central charge analysis using the monodromy technique in \cite{Malvimat:2017yaj} along the lines of \cite{Headrick:2010zt,Hartman:2013mia,Fitzpatrick:2014vua}. However a bulk proof of the conjecture similar to \cite{Faulkner:2013yia} and \cite{Lewkowycz:2013nqa} is a non trivial outstanding issue. Following this it was possible to obtain the holographic entanglement negativity for mixed state configurations of adjacent subsystems in dual $CFT$s \cite{Jain:2017sct,Jain:2017xsu,Jain:2018bai} and its covariant generalization in \cite{Jain:2017uhe}.

Very recently in \cite{Malvimat:2018txq} we have obtained the holographic entanglement negativity for the interesting mixed state configuration of disjoint intervals in proximity in a $CFT_{1+1}$ dual to static bulk $AdS_3$ geometries. In this case the replica technique requires the computation of a four point twist correlator in the $CFT_{1+1}$ leading to the entanglement negativity which involves a non universal function of the conformal cross ratio. Utilizing a monodromy technique it was possible to extract the universal part of the four point twist correlator mentioned above in the large central charge limit for the $ t $-channel ($ x \to 1 $) where $x$ was the four point conformal cross ratio. The large $c$ results for the entanglement negativity valid for the $ t $-channel could then be expressed in terms of a specific algebraic sum of the lengths of bulk space like geodesics anchored on specific combinations of the appropriate intervals. The resulting holographic construction for the entanglement negativity exhibited universal features consistent with those described in \cite{Jain:2017sct} in the adjacent interval limit.

In this article we establish a covariant holographic entanglement negativity construction for mixed state configurations of such disjoint intervals in a $CFT_{1+1}$ dual to non static bulk $AdS_3$ geometries through a large central charge analysis. Our construction is a sequel to that described in \cite{Malvimat:2018txq} and provides a generalization to characterize the entanglement for the mixed states in question in a holographic $CFT_{1+1}$ dual to non static bulk $AdS_3$ configurations. The holographic entanglement negativity in this case involves a specific algebraic sum of the lengths of bulk extremal curves anchored on appropriate intervals and their combinations, which may be expressed in terms of an algebraic sum of appropriate holographic mutual informations. Our proposal is then utilized to obtain the holographic entanglement negativity for such mixed state configurations in a $CFT_{1+1}$ with a conserved charge, dual to bulk non extremal and extremal rotating BTZ black holes. As a consistency check for our construction we subsequently compute the entanglement negativity for such mixed states in these specific holographic $CFT_{1+1}$s through a replica technique in the large central charge limit utilizing a monodromy analysis.
Quite significantly the results obtained from our holographic construction match exactly with the corresponding replica technique results obtained in the large central charge limit. Interestingly in the limit of adjacent intervals our results reproduce those described in \cite{Jain:2017uhe} and constitutes an additional substantiation of our construction. Note that the examples described above refers to $CFT_{1+1}$s dual to bulk stationary $AdS_3$ geometries. In order to elucidate the full implications of our holographic construction we subsequently obtain the time dependent holographic entanglement negativity for the mixed state under consideration in a $CFT_{1+1}$ dual to a non stationary (time dependent) bulk Vaidya-$AdS_3$ geometry whose dual $CFT$ is discussed in \cite{Anous:2017tza}. Very significantly for this example we demonstrate that the holographic entanglement negativity decreases monotonically with time illustrating the thermalization of the mixed state which corresponds to the formation of a black hole in the bulk.

This article is organized as follows. In section \ref{sec_hrt_review} we briefly review the HRT conjecture for the entanglement entropy. In section \ref{sec_chen_dj} we describe our covariant holographic entanglement negativity construction. Subsequently in section \ref{sec_chen_time_indep} we utilize our proposal to obtain the holographic entanglement negativity for mixed state configurations of disjoint intervals in $CFT_{1+1}$s dual to bulk non extremal and extremal rotating BTZ black holes.
In section \ref{sec_hen_cft_time_indep}, we obtain the entanglement negativity for the mixed states described above in the specific dual $CFT_{1+1}$s through a replica technique in the large central charge limit and compare these with the results obtained using our holographic construction. Following this in section \ref{sec_chen_time_dep}, we apply our proposal to obtain the holographic entanglement negativity of two disjoint intervals in a $CFT_{1+1}$ dual to a time dependent bulk Vaidya-$AdS_3$ configuration. Finally in section \ref{sec_summary} we present a summary of our results and conclusions.

\section{Review of the HRT conjecture}
\label{sec_hrt_review}

We begin by briefly reviewing the HRT conjecture as described in\cite{Hubeny:2007xt}. Their elegant prescription for the covariant holographic entanglement entropy involves the light sheet construction proposed by Bousso \cite{Bousso:1999xy,Bousso:1999cb,Bousso:2002ju}. In this construction the congruence of null geodesics with non positive definite expansion specifies a light sheet $ L_{ {\cal S} } $ for a space like surface $ {\cal S} $ of co dimension two in a space time manifold $ {\cal M} $. A covariant upper bound (Bousso entropy bound) for the entropy flux $ S_{ L_{\cal S} } $ (in Planck units) passing through the light sheets of the space like surface $ {\cal S} $ above is described by its area as follows
\begin{equation}
\label{eq_bousso_ee_bound}
S_{L_{\cal S}} \leq \frac{ \operatorname{Area} \left ( {\cal S} \right ) }{ 4 G_N } .
\end{equation}

The HRT conjecture states that the equality sign in the above relation holds for the holographic entanglement entropy of a subsystem in a $CFT_d$ dual to a bulk $AdS_{d+1}$ configuration. In this context the $ d $ dimensional boundary of $AdS_{d+1}$ is divided into two space like regions $ A_t $ and its complement $ A_t^c $ at a fixed time $ t $. The resulting $ (d-2) $ dimensional space like surface serving as the common boundary to these two space like regions is denoted by $ \partial A_t $.
The future and the past light sheets $ \partial L_{\pm} $ for this surface, when extended into the bulk as $ L_{\pm} $, serve as the light sheets of a $(d-1)$ dimensional bulk space like surface $ {\cal Y}_{A_t} = L_{+} \cap L_{-} $ anchored on the subsystem $ A_t $. It was proposed that amongst all such surfaces $ {\cal Y}_{A_t} $, the one with the minimal area, denoted by $ {\cal Y}_{A_t}^{min} $, determines the holographic entanglement entropy of the subsystem $ A_t $.
It was further shown that this minimal surface $ \left( {\cal Y}_{A_t}^{min} \right) $ is the extremal surface $ \left( {\cal Y}_{A_t}^{ext} \right) $ anchored on the subsystem $ A_t $. Interestingly the surface $ \left( {\cal Y}_{A_t}^{ext} \right) $ has vanishing null expansions, which explains the saturation of the covariant Bousso bound. The holographic entanglement entropy for the subsystem $ A_t $ is hence given by
\begin{equation}
\label{eq_hrt_ee}
S_{A_t}
= \frac{
\operatorname{Area} \left ( {\cal Y}_{A_t}^{min} \right )
}{
4 G_N^{(d+1)} }
= \frac{
\operatorname{Area} \left ( {\cal Y}_{A_t}^{ext} \right )
}{
4 G_N^{(d+1)} } .
\end{equation}

In this regard we note that in the context of the $AdS_3/CFT_2$ scenario, the covariant holographic entanglement entropy will involve the lengths of space like bulk extremal curves (geodesics) anchored on the corresponding spatial intervals. This concludes our brief review of the HRT construction.

\section{Covariant holographic entanglement negativity for two disjoint subsystems}
\label{sec_chen_dj}

In this section we propose our covariant holographic entanglement negativity prescription for mixed state configurations of disjoint intervals in a $CFT_{1+1}$ dual to a non static bulk $AdS_3$ geometry. To this end we first briefly outline our earlier holographic construction for the entanglement negativity of such mixed state configurations in a $CFT_{1+1}$ dual to a static bulk $AdS_3$ configuration as described in \cite{Malvimat:2018txq}.
%
%
\begin{figure}[H]
\begin{center}
\includegraphics[scale=1.5]{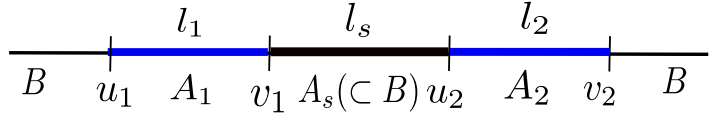}
\caption{Two disjoint intervals $ A_1 $ and $ A_2 $, separated by $ A_s $ in a $CFT_{1+1}$. The remainder of the system denoted by $B$.}
\label{fig_en_dj_int}
\end{center}
\end{figure}
%
%
The entanglement negativity for the mixed state of disjoint intervals $A_1\in\left [ u_1, v_1 \right ] $ (of length $ l_1 $) and $A_2\in\left [ u_2, v_2 \right ] $ (of length $ l_2 $), separated by the interval $A_S\in\left [ v_1, u_2 \right ] $ (of length $ l_s $) in the dual $CFT_{1+1}$ as illustrated in figure \ref{fig_en_dj_int}, may be obtained through a replica technique involving a specific four point twist correlator as follows
\begin{equation}
\label{eq_ent_neg_ito_twist_correlator}
{\cal E} = \lim_{ n_e \to 1 } \ln
\left \langle {\cal T}_{n_e} (z_1)
{\overline{\cal T}}_{n_e} (z_2)
{\overline{\cal T}}_{n_e} (z_3)
{\cal T}_{n_e} (z_4) \right \rangle_{\mathbb{C}} .
\end{equation}
Here we have identified $ u_1 \equiv z_1, v_1 \equiv z_2, u_2 \equiv z_3, v_2 \equiv z_4 $.
The limit $ n_e \to 1 $ is taken through an analytic continuation of even sequences $ \{ n_e \} $ to $ n_e = 1 $.

The four point twist correlator in eq.\@ \eqref{eq_ent_neg_ito_twist_correlator} is in general non universal and depends on the full operator content of the underlying theory as described in \cite{Calabrese:2012nk}. However, for two disjoint intervals in proximity (described by the condition $ 1/2 < x < 1 $), in the large central charge limit, the following universal form may be obtained through the monodromy technique 
\cite{Hartman:2013mia,Kulaxizi:2014nma,Malvimat:2017yaj}
\begin{equation}
\label{eq_twist_4pt_fn_ito_x}
\lim_{ n_e \to 1 } \left \langle {\cal T}_{n_e} (z_1)
{\overline{\cal T}}_{n_e} (z_2)
{\overline{\cal T}}_{n_e} (z_3)
{\cal T}_{n_e} (z_4) \right \rangle_{ \mathbb{C} }
= \left ( 1 - x \right )^{ 2 \hat h } .
\end{equation}
Here the four point cross ratio $ x $ is given by $ ( z_{12} z_{34} ) / ( z_{13} z_{24} ) $ with $ z_{ij} \equiv z_i - z_j $, and $ \hat h $ is the conformal dimension of the operator with the dominant contribution in the corresponding conformal block expansion. The entanglement negativity may then be obtained from eq.\@ \eqref{eq_ent_neg_ito_twist_correlator} as \cite{Malvimat:2018txq}
\begin{equation}
\label{eq_ent_neg_expr_dj_int}
{\cal E} = \frac{ c }{ 4 }
\ln \left (
\frac{ |z_{13}| |z_{24}| }
{ |z_{14}| |z_{23}| }
\right ) .
\end{equation}
The two point twist correlator in a holographic $CFT_{1+1}$ is given as
\begin{equation}
\label{eq_twist_2pt_fn_cft}
\left \langle {\cal T}_{n_e} (z_i)
\overline{{\cal T}}_{n_e} (z_j) \right \rangle_{\mathbb{C}}
\sim \left | z_{ij} \right |^{- 2 \Delta_{ {\cal T}_{n_e}} } .
\end{equation}
Using the $AdS_3/CFT_2$ dictionary this two point twist correlator (in the geodesic approximation) may be described in terms of the length $ {\cal L}_{ij} $ of the space like geodesic in the bulk, anchored on the corresponding interval in the boundary as follows \cite{Ryu:2006ef}
\begin{equation}
\label{eq_twist_2pt_fn_ads_cft}
\left \langle {\cal T}_{n_e} (z_i)
\overline{{\cal T}}_{n_e} (z_j) \right \rangle_{\mathbb{C}}
\sim
\exp \left ( - \frac{ \Delta_{{\cal T}_{n_e}} {\cal L}_{ij} }{ R } \right ) ,
\end{equation}
with $ R $ being the $AdS_3$ length scale.\\
Utilizing eqs.\@ \eqref{eq_twist_2pt_fn_cft} and \eqref{eq_twist_2pt_fn_ads_cft},
it is now possible to express the four point twist correlator given in eq.\@ \eqref{eq_twist_4pt_fn_ito_x} as follows
\begin{equation}
\label{eq_twist_4pt_fn_ito_geo_len}
\begin{aligned}
& \lim_{ n_e \to 1 }
\left \langle {\cal T}_{n_e} (z_1)
\overline{{\cal T}}_{n_e} (z_2)
\overline{{\cal T}}_{n_e} (z_3)
{\cal T}_{n_e} (z_4) \right \rangle_{\mathbb{C}} \\
& = \exp \left [ \frac{ c }{ 8 R }
\left ( {\cal L}_{13} + {\cal L}_{24}
- {\cal L}_{14} - {\cal L}_{23} \right ) \right ] .
\end{aligned}
\end{equation}
From the expression for the four point twist correlator in eq.\@ \eqref{eq_twist_4pt_fn_ito_geo_len} and using eq.\@ \eqref{eq_ent_neg_ito_twist_correlator} 
the holographic entanglement negativity for the mixed state configuration of the two disjoint intervals in proximity may be expressed as
\begin{equation}
\label{eq_hen_dj_geo_len}
{\cal E} = \frac{ 3 } { 16 G_N^{(3)} }
\left ( {\cal L}_{ A_1 \cup A_s } + {\cal L}_{ A_2 \cup A_s }
- {\cal L}_{ A_1 \cup A_2 \cup A_s } - {\cal L}_{ A_s } \right ) .
\end{equation}
Here the Brown-Henneaux formula $ c = \frac{ 3 R }{ 2 G_N^{(3)} } $ \cite{Brown:1986nw} has been used.

According to the HRT conjecture the computation of holographic entanglement entropy in the time dependent scenario involves the area of the corresponding co dimension two bulk extremal surface anchored on the respective subsystem in the boundary. In the context of the $AdS_3/CFT_2$ scenario these are extremal curves anchored on the corresponding intervals. This motivates us to propose the following covariant construction for the time dependent holographic entanglement negativity of the mixed state configuration of two disjoint intervals in proximity in the $AdS_3/CFT_2$ framework as
\begin{equation}
\label{eq_chen_dj_geo_len}
{\cal E} = \frac{ 3 } { 16 G_N^{(3)} }
\left ( {\cal L}_{ A_1^t \cup A_s^t }^{ext}
+ {\cal L}_{ A_2^t \cup A_s^t }^{ext}
- {\cal L}_{ A_1^t \cup A_2^t \cup A_s^t }^{ext}
- {\cal L}_{ A_s^t }^{ext}
\right ) ,
\end{equation}
where $ {\cal L}_{\gamma}^{ext} $ denotes the length of the bulk space like extremal curve anchored on the subsystem ($ \gamma $) in the dual $CFT_{1+1}$.
Note that here $ A_i^t $ denotes the subsystem $ A_i $ at a fixed time $ t $ as described earlier in section \ref{sec_hrt_review}. The HRT conjecture described in eq.\@ \eqref{eq_hrt_ee} may now be utilized to arrive at the following
\begin{equation}
\label{eq_chen_dj_hee}
{\cal E} = \frac{ 3 } {4}
\left ( S_{ A_1^t \cup A_s^t } + S_{ A_s^t \cup A_2^t }
- S_{ A_1^t \cup A_2^t \cup A_s^t } - S_{ A_s^t } \right ) .
\end{equation}
The covariant holographic entanglement negativity described in eq.\@ \eqref{eq_chen_dj_hee} may be further expressed in terms of a specific sum of the holographic mutual informations between appropriate combinations of the subsystems as follows\footnote{This interesting connection in the holographic limit between the universal parts of the entanglement negativity and the mutual information which are distinct quantities in quantum information theory has been reported in the literature. In particular for two adjacent intervals, they have been shown to be identical \cite{Coser:2014gsa,Wen:2015qwa}.}
\begin{equation}
\label{eq_chen_dj_mut_info}
{\cal E} = \frac{ 3 } {4}
\left [ {\cal I} ( A_1^t \cup A_s^t, A_2^t )
- {\cal I} ( A_s^t, A_2^t ) \right ] .
\end{equation}
Here the holographic mutual information between the subsystems $ A_i^t $ and $ A_j^t $
is given by
$ {\cal I}(A_i^t,A_j^t) \equiv S_{A_i^t} + S_{A_j^t} - S_{A_i^t \cup A_j^t} $. Interestingly in the limit $ A_s^t \to \emptyset $, we recover the exact expression as reported in \cite{Jain:2017uhe} for the the mixed state holographic entanglement negativity of adjacent intervals in terms of the holographic mutual information between the subsystems $ A_1^t $ and $ A_2^t $.

\section{Covariant holographic entanglement negativity in the dual $CFT_{1+1}$ }
\label{sec_chen_time_indep}

In this section as a consistency check we utilize our covariant holographic construction as described in eq.\@ \eqref{eq_chen_dj_geo_len} above, to compute the entanglement negativity for mixed states of two disjoint intervals in proximity in a $CFT_{1+1}$ with a conserved charge, dual to bulk stationary $AdS_3$ geometries described by non extremal and extremal rotating BTZ black holes.

\subsection{Non extremal BTZ black hole}
\label{subsec_chen_non_ext_btz}

We first focus on the $CFT_{1+1}$ with a conserved charge dual to the non extremal rotating BTZ (black string) whose metric (with the $AdS_3$ radius $ R $ set to unity) is given as 
\begin{equation}
\label{eq_metric_non_ext_btz}
\begin{aligned}
ds^2 & = 
- \frac{
\left ( r^2 - r_{+}^2 \right ) \left ( r^2 - r_{-}^2 \right )
}{ r^2 } dt^2 \\
& + \frac{ r^2 }
{
\left ( r^2 - r_{+}^2 \right ) \left ( r^2 - r_{-}^2 \right )
}
dr^2 + r^2 \left ( d\phi - \frac{ r_{+} r_{-} }{ r^2 } dt \right )^2 ,
\end{aligned}
\end{equation}
where the coordinate $ \phi $ is non compact and $ r_{\pm} $ denote the radii of the inner and the outer horizons respectively. The mass $ M $, angular momentum $ J $, Hawking temperature $ T_H $, and the angular velocity $ \Omega $ may be expressed in terms of the horizon radii $ r_{\pm} $ as follows
\begin{equation}
\label{eq_relations_non_ext_btz}
\begin{alignedat}{2}
M & = r_{+}^2 + r_{-}^2 ,
& \qquad \qquad \qquad \qquad
J & = 2 r_{+} r_{-} , \\
T_H & = \frac{ 1 }{ \beta }
= \frac{ r_{+}^2 - r_{-}^2 }{ 2 \pi r_{+} } ,
& \Omega & = \frac{ r_{-} }{ r_{+} } .
\end{alignedat}
\end{equation}
The finite temperature dual $CFT_{1+1}$ with the bulk angular momentum $ J $ as the conserved charge may be defined on a twisted infinite cylinder.\footnote{\label{fnote_twisted_cyl}This is evident from the conformal map described in eq.\@ \eqref{eq_non_ext_btz_large_r_coord_trfn}. See \cite{Caputa:2013eka} for details.} The temperatures relevant to the left and the right moving modes respectively in the dual $CFT$ are given as
\begin{equation}
\label{eq_non_ext_btz_temp_relns}
T_{\pm} = \frac{ 1 }{ \beta_{\pm} }
= \frac{ r_{+} \mp r_{-} }{ 2 \pi } .
\end{equation}
To obtain the lengths of the extremal curves (which are space like geodesics) in this bulk geometry we map the BTZ metric in eq.\@ \eqref{eq_metric_non_ext_btz} to the Poincar\'{e} patch of the $AdS_3$ space time as follows
\begin{equation}
\label{eq_non_ext_btz_coord_trfn}
\begin{gathered}
w_{\pm} = \sqrt{ \frac{ r^2 - r_{+}^2 }{ r^2 - r_{-}^2 } }
\exp \left ( 2 \pi T_{\pm} u_{\pm} \right )
\equiv X \pm T , \\
Z = \sqrt{ \frac{ r_{+}^2 - r_{-}^2 }{ r^2 - r_{-}^2 } }
\exp \left ( \pi T_{+} u_{+} + \pi T_{-} u_{-} \right ) ,
\end{gathered}
\end{equation}
where $ u_{\pm} = \phi \pm t $. The resulting $AdS_3$ metric is then given by
\begin{equation}
\label{eq_metric_ads3_poincare}
ds^2 = \frac{ dw_{+} dw_{-} + dZ^2 }{ Z^2 }
\equiv \frac{ - dT^2 + dX^2 + dZ^2 }{ Z^2 } .
\end{equation}
The length of the space like geodesic in this bulk geometry, which is anchored on an interval $ \gamma $ of length $ \Delta \phi = \left | \phi_1 - \phi_2 \right | $ in the dual $CFT_{1+1}$, may now be obtained as follows \cite{Hubeny:2007xt}
\begin{align}
\label{eq_non_ext_btz_geo_len}
{\cal L}_{\gamma} & = \ln \left [ \frac{
\left ( \Delta \phi \right )^2
}{
\varepsilon_{i} \varepsilon_{j} } \right ] ,
& \varepsilon_{i} & = \sqrt{ \frac{ r_{+}^2 - r_{-}^2 }{ r_{\infty}^2 } }
\exp \left ( r_{+} \phi_{i} - t_0 r_{-} \right ) .
\end{align}
Here $ r_{\infty} $ and $ \varepsilon_{i} $ ($ i, j = 1, 2 $) denote the infra red cut of{}fs for the bulk in the BTZ and the Poincar\'{e} coordinates respectively. Using the above expressions in eq.\@ \eqref{eq_non_ext_btz_geo_len} the geodesic length $ {\cal L}_{\gamma} $ may now be expressed in terms of the original BTZ coordinates as follows \cite{Hubeny:2007xt}
\begin{equation}
\label{eq_non_ext_btz_geo_len_org_coord}
{\cal L}_{\gamma}
= \ln \left [ 
\frac{ \beta_{+} \beta_{-} }{ \pi^2 a^2 }
\sinh \left ( \frac{ \pi \Delta \phi }{ \beta_{+} } \right )
\sinh \left ( \frac{ \pi \Delta \phi }{ \beta_{-} } \right )
\right ] .
\end{equation}
Here $ a $ is the UV cut of{}f of the $CFT_{1+1}$ in the boundary which is related to the bulk infra red cut of{}f $ {r_\infty} $ through the $AdS/CFT$ dictionary by $ a = 1 / {r_\infty} $.

The covariant holographic entanglement negativity for the mixed state configuration of disjoint intervals in proximity in the $CFT_{1+1}$ with a conserved charge, dual to the non extremal BTZ black hole in the bulk, may now be obtained by utilizing our covariant holographic negativity construction described by eq.\@ \eqref{eq_chen_dj_geo_len} as follows
\begin{equation}
\label{eq_non_ext_btz_chen}
\begin{aligned}
{\cal E}
& = \frac{ c }{ 8 } \ln \left [ \frac
{
\sinh \left \{ \frac{ \pi \left ( l_1 + l_s \right ) }{ \beta_{+} } \right \}
\sinh \left \{ \frac{ \pi \left ( l_2 + l_s \right ) }{ \beta_{+} } \right \}
}{
\sinh \left ( \frac{ \pi l_s }{ \beta_{+} } \right )
\sinh \left \{ \frac{ \pi \left ( l_1 + l_2 + l_s \right ) }{ \beta_{+} } \right \}
} \right ] \\
& \qquad + \frac{ c }{ 8 } \ln \left [ \frac
{
\sinh \left \{ \frac{ \pi \left ( l_1 + l_s \right ) }{ \beta_{-} } \right \}
\sinh \left \{ \frac{ \pi \left ( l_2 + l_s \right ) }{ \beta_{-} } \right \}
}{
\sinh \left ( \frac{ \pi l_s }{ \beta_{-} } \right )
\sinh \left \{ \frac{ \pi \left ( l_1 + l_2 + l_s \right ) }{ \beta_{-} } \right \}
} \right ] \\
& = \frac{ 1 }{ 2 } \left ( {\cal E}_L + {\cal E}_R \right ) .
\end{aligned}
\end{equation}
Here $ l_1 $ and $ l_2 $ denote the respective lengths of the disjoint intervals and their separation is given by $ l_s $ and $L/R$ denotes the left and the right moving sector in the dual $CFT_{1+1}$. Note that the covariant holographic entanglement negativity in this case decouples into $ {\cal E}_L $ and $ {\cal E}_R $ with the left and right moving inverse temperatures $ \beta_{+} $ and $ \beta_{-} $ respectively \cite{2014NJPh...16l3020E}. It is interesting to note that the above expression for entanglement negativity for the mixed state configuration of disjoint intervals is independent of the UV cut of{}f $ a $. When the intervals in question are adjacent (obtained through the limit $ l_s \to a $), the above expression in eq.\@ \eqref{eq_non_ext_btz_chen} exactly reproduces the corresponding entanglement negativity result in \cite{Jain:2017uhe}.

\subsection{Extremal BTZ black hole}
\label{subsec_chen_ext_btz}

Having computed the holographic entanglement negativity for the mixed state of disjoint intervals in a $CFT_{1+1}$ dual to a bulk non extremal rotating BTZ black hole, we now turn our attention to the corresponding case in a $CFT_{1+1}$ dual to an an extremal rotating BTZ black hole. The metric for the dual bulk extremal black hole in the BTZ coordinates is given by

\begin{equation}
\label{eq_metric_ext_btz}
\begin{aligned}
ds^2 & = - \frac{ \left ( r^2 - r_{0}^2 \right )^2 }{ r^2 } dt^2
+ \frac{ r^2 }{ \left ( r^2 - r_{0}^2 \right )^2 } dr^2 \\
& \qquad + r^2 \left ( d\phi - \frac{ r_{0}^2 }{ r^2 } dt \right )^2 .
\end{aligned}
\end{equation}
In this extremal limit described by $ r_{+} = r_{-} = r_0 $, the Hawking temperature vanishes $ \left ( T_H = 0 \right ) $, whereas the mass and the angular momentum coincide 
$ \left ( M = J = 2 r_0^2 \right ) $.

To compute the bulk space like geodesic lengths $ {\cal L}_{\gamma} $ as earlier, we first consider the coordinate transformation mapping the extremal BTZ metric in eq.\@ \eqref{eq_metric_ext_btz} to the $AdS_3$ metric in the Poincar\'{e} coordinates given in eq.\@ \eqref{eq_metric_ads3_poincare}
as follows
\begin{equation}
\label{eq_ext_btz_coord_trfn}
\begin{gathered}
\begin{alignedat}{2}
w_{+} & = \phi + t - \frac{ r_0 }{ r^2 - r_0^2 } ,
& \qquad
w_{-} & = \frac{ 1 }{ 2 r_0 }
\exp \left [ 2 r_0 \left ( \phi - t \right ) \right ] ,
\end{alignedat}
\\
Z = \frac{ 1 }{ \sqrt{ r^2 - r_0^2 } }
\exp \left [ r_0 \left ( \phi - t \right ) \right ] .
\end{gathered}
\end{equation}
The length $ {\cal L}_{\gamma} $ of the space like bulk geodesic anchored on the interval 
$ \gamma $ of length $ \Delta \phi $ may then be obtained as \cite{Hubeny:2007xt}
\begin{align}
\label{eq_ext_btz_geo_len}
{\cal L}_{\gamma} & = \ln \left [ \frac{
\left ( \Delta \phi \right )^2 }
{ \varepsilon_{i} \varepsilon_{j} } \right ] ,
& \varepsilon_{i} & = \frac{ 1 }{ r_{\infty} }
\exp \left [ r_0 \left ( \phi_{i} - t_0 \right ) \right ] ,
\end{align}
where $ i, j = \{ 1, 2 \} $. Finally, the above expression for the space like bulk geodesic length ${\cal L}_{\gamma}$
in eq.\@ \eqref{eq_ext_btz_geo_len} may be expressed in the original BTZ coordinates as follows
\begin{equation}
\label{eq_ext_btz_geo_len_org_coord}
{\cal L}_{\gamma} = \ln \left ( \frac{ \Delta \phi }{ a } \right )
+ \ln \left [ \frac{ 1 }{ r_0 a }
\sinh \left ( r_0 \Delta \phi \right ) \right ] .
\end{equation}
We now utilize our covariant construction described in eq.\@ \eqref{eq_chen_dj_geo_len} to obtain the holographic entanglement negativity for the mixed state configuration of disjoint intervals in proximity in the $CFT_{1+1}$ under consideration as 
follows
\begin{equation}
\label{eq_ext_btz_chen}
\begin{aligned}
{\cal E} & = \frac{ c }{ 8 }
\ln \left [ \frac{
\left ( l_1 + l_s \right )
\left ( l_2 + l_s \right )
}{
l_s \left ( l_1 + l_2 + l_s \right )
} \right ] \\
& \quad + \frac{ c }{ 8 }
\ln \left [ \frac
{
\sinh \left \{ r_0 \left ( l_1 + l_s \right ) \right \}
\sinh \left \{ r_0 \left ( l_2 + l_s \right ) \right \}
}{
\sinh \left ( r_0 l_s \right )
\sinh \left \{ r_0 \left ( l_1 + l_2 + l_s \right ) \right \}
}
\right ] .
\end{aligned}
\end{equation}
Once again we note that this result is independent of the UV cut of{}f $a$. Interestingly the holographic entanglement negativity in the above expression decouples into two independent parts with distinct significances. The first term is exactly half of the corresponding holographic entanglement negativity for the zero temperature mixed state configuration of two disjoint intervals in proximity described in \cite{Malvimat:2018txq}.
The second term on the other hand resembles the finite temperature entanglement negativity of the mixed state configuration in question, at the Frolov-Throne temperature $ T_{ FT } = \frac{ r_0 }{ \pi } $ \cite{Frolov:1989jh,Caputa:2013lfa}. As earlier in the adjacent limit ($ l_s \to a $), our result in eq.\@ \eqref{eq_ext_btz_chen} exactly reproduces the corresponding holographic entanglement negativity of the mixed state configuration of adjacent intervals described in\cite{Jain:2017uhe}.

\section{Entanglement negativity for disjoint intervals in the dual $CFT_{1+1}$}
\label{sec_hen_cft_time_indep}

In this section we compute the entanglement negativity for the mixed state configurations of disjoint intervals in proximity in $CFT_{1+1}$s dual to non extremal and extremal rotating BTZ black holes, through a replica technique.
We then compare these results with the corresponding holographic results obtained in the previous section. It is observed that they match exactly in the large central charge limit and serves as a strong consistency check for our holographic construction. For this purpose we briefly recapitulate the essential elements of the computation of entanglement negativity for such mixed states in a generic $CFT_{1+1}$ described in \cite{Calabrese:2012ew,Calabrese:2012nk,Malvimat:2018txq}.

We begin by noting that for a $CFT_{1+1}$ on the complex plane, a universal form for the four point twist correlator in the large central charge limit has been obtained in \cite{Malvimat:2017yaj}, as described earlier in eq.\@ \eqref{eq_twist_4pt_fn_ito_x}. On the other hand $CFT_{1+1}$s with a conserved charge dual to bulk non extremal and the extremal rotating BTZ black holes must be defined on twisted infinite cylinders (see footnote \ref{fnote_twisted_cyl}).
Hence the four point twist correlator in eq.\@ \eqref{eq_twist_4pt_fn_ito_x} for such $CFT_{1+1}$s are computed on such twisted cylinders. The corresponding conformal maps from the complex plane to the twisted cylinder are described in the following subsections. The transformation property of the four point twist correlator in eq.\@ \eqref{eq_twist_4pt_fn_ito_x} under a conformal map $ z \to w $ is given as follows
\begin{equation}
\label{eq_twist_4pt_fn_trfn}
\begin{aligned}
& \left \langle {\cal T}_{n_e} (w_1)
\overline{{\cal T}}_{n_e} (w_2)
\overline{{\cal T}}_{n_e} (w_3)
{\cal T}_{n_e} (w_4) \right \rangle_{cyl} \\
& = \prod_{i=1}^{4}
\left [ \left ( \frac{ d w(z) }{ d z } \right )^{ - \Delta_i }
\right ]_{ z = z_i } \\
& \qquad \qquad \times \left \langle {\cal T}_{n_e} (z_1)
\overline{{\cal T}}_{n_e} (z_2)
\overline{{\cal T}}_{n_e} (z_3)
{\cal T}_{n_e} (z_4) \right \rangle_{ \mathbb{C} } ,
\end{aligned}
\end{equation}
where $ \Delta_i $ are the scaling dimensions of the twist fields at the locations $ z = z_i $. Here $ z $ and $ w $ are the coordinates on the plane and the twisted cylinder respectively. 
The entanglement negativity for the mixed state of the two disjoint intervals described above may then be computed by utilizing eqs.\@ \eqref{eq_ent_neg_ito_twist_correlator}, \eqref{eq_twist_4pt_fn_ito_x} and \eqref{eq_twist_4pt_fn_trfn} by identifying the coordinates ($ w_i $) on the twisted cylinder as
$ w_1 \equiv u_1, w_2 \equiv v_1, w_3 \equiv u_2, w_4 \equiv v_2 $, as described below.

\subsection{Non extremal BTZ black hole}
\label{subsec_ent_neg_non_ext_btz}

We now proceed to compute the entanglement negativity for the mixed state configuration of disjoint intervals in proximity in the finite temperature $CFT_{1+1}$ dual to a non extremal rotating BTZ black hole. To this end we consider the partition function for this $CFT$, which is described as follows \cite{Hubeny:2007xt,Caputa:2013lfa}
\begin{equation}
\label{eq_non_ext_partition_fn}
\begin{aligned}
Z \left ( \beta \right ) & = \operatorname{Tr}
\exp \left [ - \beta \left ( H - i \Omega_{E} J \right ) \right ] \\
& = \operatorname{Tr} \exp \left ( - \beta_{+} L_{0} - \beta_{-} \bar{L}_{0} \right ) .
\end{aligned}
\end{equation}
Here $\rho = \exp \left [ - \beta \left ( H - i \Omega_{E} J \right ) \right ] $ is the appropriate density matrix, and the angular momentum which is a conserved charge is denoted by $ J $. $ \Omega_{E} = - i \Omega $ is the Euclidean angular velocity and the Hamiltonian is given by $ H = L_{0} + \bar{L}_{0} $ where $ L_{0} $ and $ \bar{L}_{0} $ are the standard Virasoro zero modes. As described earlier in subsection \ref{subsec_chen_non_ext_btz}, the non extremal BTZ metric given in eq.\@ \eqref{eq_metric_non_ext_btz} is mapped to the Poincar\'{e} patch of $AdS_3$ by the coordinate transformations described in eq.\@ \eqref{eq_non_ext_btz_coord_trfn}. In the limit $ r \gg r_{\pm} $, the conformal map from the $CFT_{1+1}$ on the complex plane to that on the twisted infinite cylinder reduces to
\begin{equation}
\label{eq_non_ext_btz_large_r_coord_trfn}
\begin{gathered}
w(z) \equiv
\exp \left [ \frac{ 2 \pi }{ \beta_{+} } \left ( \phi + i \tau \right ) \right ]
= \exp \left ( \frac{ 2 \pi }{ \beta_{+} } z \right ) , \\
\bar{w}(\bar{z}) \equiv
\exp \left [ \frac{ 2 \pi }{ \beta_{-} } \left ( \phi - i \tau \right ) \right ]
= \exp \left ( \frac{ 2 \pi }{ \beta_{-} } \bar{z} \right ) .
\end{gathered}
\end{equation}
Utilizing eqs.\@ \eqref{eq_ent_neg_ito_twist_correlator}, \eqref{eq_twist_4pt_fn_ito_x}, \eqref{eq_twist_4pt_fn_trfn} and \eqref{eq_non_ext_btz_large_r_coord_trfn}, the entanglement negativity of the mixed state in question in the large central charge limit may then be obtained as follows
\begin{equation}
\label{eq_non_ext_btz_ent_neg}
\begin{aligned}
{\cal E}
& = \frac{ c }{ 8 } \ln \left [ \frac
{
\sinh \left \{ \frac{ \pi \left ( l_1 + l_s \right ) }{ \beta_{+} } \right \}
\sinh \left \{ \frac{ \pi \left ( l_2 + l_s \right ) }{ \beta_{+} } \right \}
}{
\sinh \left ( \frac{ \pi l_s }{ \beta_{+} } \right )
\sinh \left \{ \frac{ \pi \left ( l_1 + l_2 + l_s \right ) }{ \beta_{+} } \right \}
} \right ] \\
& \qquad + \frac{ c }{ 8 } \ln \left [ \frac
{
\sinh \left \{ \frac{ \pi \left ( l_1 + l_s \right ) }{ \beta_{-} } \right \}
\sinh \left \{ \frac{ \pi \left ( l_2 + l_s \right ) }{ \beta_{-} } \right \}
}{
\sinh \left ( \frac{ \pi l_s }{ \beta_{-} } \right )
\sinh \left \{ \frac{ \pi \left ( l_1 + l_2 + l_s \right ) }{ \beta_{-} } \right \}
} \right ] \\
& = \frac{ 1 }{ 2 } \left ( {\cal E}_L + {\cal E}_R \right ) .
\end{aligned}
\end{equation}
Note that the universal part of the entanglement negativity for the $CFT_{1+1}$ decouples into $ {\cal E}_{L} $ and $ {\cal E}_{R} $ components depending only on the left and right moving inverse temperatures $ \beta_{+} $ and $ \beta_{-} $ respectively \cite{2014NJPh...16l3020E}
(see also the discussion after eq.\@ \eqref{eq_non_ext_btz_chen}).
This decoupling corresponds to the two mutually commuting chiral components of the Hamiltonian $ H = L_{0} + \bar{L}_{0} $ as described in \cite{2014NJPh...16l3020E}. Significantly, the entanglement negativity for the mixed state configuration of disjoint intervals obtained in the large central charge limit exactly matches with the corresponding result in eq.\@ \eqref{eq_non_ext_btz_chen} obtained from our covariant holographic entanglement negativity construction.

\subsection{Extremal BTZ black hole}
\label{subsec_ent_neg_ext_btz}

As stated earlier in subsection \ref{subsec_chen_ext_btz}, the extremal BTZ metric given in eq.\@ \eqref{eq_metric_ext_btz} is mapped to the Poincar\'{e} patch of $AdS_3$ by the coordinate transformations described in eq.\@ \eqref{eq_ext_btz_coord_trfn}.

In the asymptotic limit ($ r \to \infty $), the relevant conformal map from the complex plane to the twisted infinite cylinder in this case may be obtained from the coordinate transformations described in eq.\@ \eqref{eq_ext_btz_coord_trfn} after a Wick rotation $ t \to i \tau $ as follows
\begin{equation}
\label{eq_ext_btz_large_r_coord_trfn}
\begin{gathered}
w(z) \equiv \phi + i \tau = z , \\
\bar{w} (\bar{z}) \equiv \frac{ 1 }{ 2 r_0 }
\exp \left [ 2 r_0 \left ( \phi - i \tau \right ) \right ]
= \frac{ 1 }{ 2 r_0 } \exp \left ( 2 r_0 \bar{z} \right ) .
\end{gathered}
\end{equation}
Utilizing eqs.\@ \eqref{eq_ent_neg_ito_twist_correlator}, \eqref{eq_twist_4pt_fn_ito_x}, \eqref{eq_twist_4pt_fn_trfn} and \eqref{eq_ext_btz_large_r_coord_trfn}, the entanglement negativity for the mixed state configuration of disjoint intervals in proximity in the large central charge limit in the $CFT_{1+1}$ dual to an extremal BTZ black hole may then be determined as follows
\begin{equation}
\label{eq_ext_btz_ent_neg}
\begin{aligned}
{\cal E} & = \frac{ c }{ 8 }
\ln \left [
\frac{
\left ( l_1 + l_s \right )
\left ( l_2 + l_s \right )
}{
l_s \left ( l_1 + l_2 + l_s \right )
} \right ] \\
& \qquad + \frac{ c }{ 8 }
\ln \left [
\frac
{
\sinh \left \{ r_0 \left ( l_1 + l_s \right ) \right \}
\sinh \left \{ r_0 \left ( l_2 + l_s \right ) \right \}
}{
\sinh \left ( r_0 l_s \right )
\sinh \left \{ r_0 \left ( l_1 + l_2 + l_s \right ) \right \}
}
\right ] .
\end{aligned}
\end{equation}
Significantly the first term in the above expression indicates the entanglement negativity of the disjoint intervals in proximity for the $CFT_{1+1}$ in its ground state while the second term describes the entanglement negativity at an ef{}fective Frolov-Thorne temperature $T_{FT}$ \cite{Frolov:1989jh,Caputa:2013lfa} (see the discussion after eq.\@ \eqref{eq_ext_btz_chen}).
Remarkably the entanglement negativity for the $CFT_{1+1}$ obtained through the replica technique in the large central charge limit matches exactly with the holographic result described in eq.\@ \eqref{eq_ext_btz_chen}. Once again this constitutes a robust consistency check for our holographic construction.

\section{Time dependent holographic entanglement negativity}
\label{sec_chen_time_dep}

In the earlier sections we have described the application of our covariant holographic entanglement negativity construction to time independent mixed state configuration of disjoint intervals in proximity in $CFT_{1+1}$s dual to stationary bulk $AdS_3$ configurations and compared our results with those obtained through a replica technique. However note that our covariant holographic construction is also valid for time dependent mixed states in $CFT_{1+1}$s dual to non stationary bulk $AdS_3$ configurations. In this section we address this significant issue of time dependent mixed state entanglement through our covariant holographic entanglement negativity construction described in section \ref{sec_chen_dj}. In this context we obtain the holographic entanglement negativity for the time dependent mixed state configuration of disjoint intervals in proximity in a $CFT_{1+1}$ dual to a bulk Vaidya-$AdS_3$ configuration. This bulk $AdS_3$ geometry characterizes one of the simplest time dependent gravitational configurations and describes the collapse process and subsequent black hole formation.

The Vaidya-$AdS_3$ metric may be described in the Poincar\'{e} coordinates
as \cite{Hubeny:2007xt,AbajoArrastia:2010yt}
\begin{equation}
\label{eq_metric_vaidya_ads}
ds^2 = - \left [ r^2 - m(v) \right ] dv^2
+ 2 dr dv + r^2 dx^2 .
\end{equation}
Here $ m(v) $ denotes the mass function for the light cone time $ v $. For a constant $ m(v) $, the above metric describes a non rotating BTZ black hole. In this scenario, the space like geodesic in the bulk Vaidya-$AdS_3$, anchored on the interval $ x \in \left [- l/2, l/2 \right ] $ (of length $ l $) in the dual $CFT_{1+1}$, may be characterized by the functions $ r(x) $ and $ v(x) $ of the boundary spatial coordinate $ x $ with the boundary conditions given as
\begin{align}
\label{eq_vaidya_ads_geodesic_boundary_cond}
r(-l/2) & = r(l/2) = r_{\infty} ,
& v(-l/2) & = v(l/2) = v .
\end{align}
Here the UV cut of{}f $ r_{\infty} = 1/\varepsilon $, where $ \varepsilon $ is the lattice spacing. The length of the space like geodesic described above may be given as
\begin{equation}
\label{eq_vaidya_ads_geo_len}
{\cal L} = \int_{-l/2}^{l/2}
\sqrt{ r^2 + 2 \dot{r} \dot{v} - f(r,v) \dot{v}^2 } dx ,
\end{equation}
where $ \dot{v} $ and $ \dot{r} $ denote the derivatives of the functions $ v $ and $ r $ with respect to the coordinate $ x $ respectively.

To extract an explicit form of the geodesic length $ {\cal L} $ from the above expression, it is required to employ the adiabatic approximation as outlined in \cite{Hubeny:2007xt} wherein the mass function $ m(v) $ is assumed to be a slowly varying function of $ v $ and $ \dot{m}(v) \equiv \frac{ dm }{ dv } / \frac{ dx }{ dv } \ll 1 $. The length ${\cal L}$ of the bulk space like geodesic may then be computed as \cite{Hubeny:2007xt}
\begin{equation}
\label{eq_vaidya_ads_geo_len_adiabatic}
{\cal L} = {\cal L}_{reg} + 2 \ln \left ( 2 r_{\infty} \right ) .
\end{equation}
Here the regularized part $ {\cal L}_{\textrm{reg}} $ of the length of the space like geodesic as mentioned in the above expression is given as
\begin{equation}
\label{eq_vaidya_ads_geo_len_reg}
{\cal L}_{reg} = \ln \left [
\frac{ \sinh^2 \left ( \sqrt{ m(v) l/2 } \right ) }{ m(v) } \right ] .
\end{equation}
We may now use the expression for $ {\cal L} $ in eq.\@ \eqref{eq_vaidya_ads_geo_len} to compute the covariant holographic entanglement negativity for the mixed state of disjoint intervals in proximity by utilizing our holographic conjecture as described in eq.\@ \eqref{eq_chen_dj_geo_len} as follows
\begin{equation}
\label{eq_vaidya_ads_chen}
\begin{aligned}
& {\cal E} = \frac{ 3 }{ 16 G_N^{(3)} } \times \\
& \ln
\left [ \frac
{
\sinh^2 \left \{ \sqrt{ m(v) \left ( l_1 + l_s \right ) / 2 } \right \}
\sinh^2 \left \{ \sqrt{ m(v) \left ( l_2 + l_s \right ) / 2 } \right \}
}{
\sinh^2 \left ( \sqrt{ m(v) l_s/2 } \right )
\sinh^2 \left \{ \sqrt{ m(v) \left ( l_1 + l_2 + l_s \right ) / 2 } \right \}
}
\right ] .
\end{aligned}
\end{equation}
Interestingly, unlike the case for adjacent intervals described in \cite{Jain:2017uhe}, for the case of disjoint intervals, the entanglement negativity has no divergent part.

\begin{figure}[H]
\begin{center}
\includegraphics[scale=1]{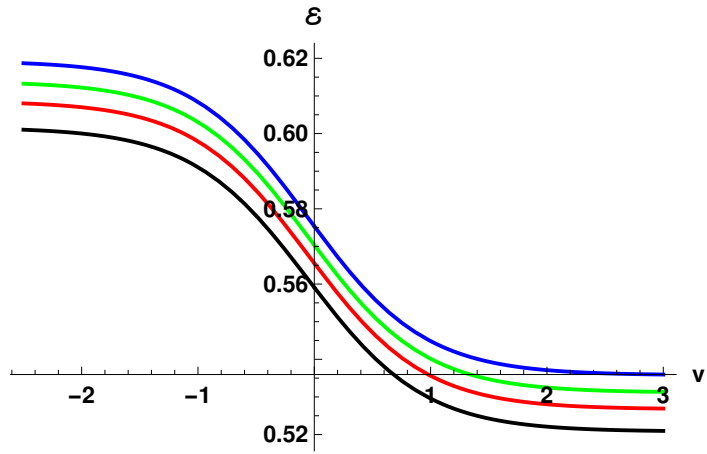}
\caption{Plots of entanglement negativity $ {\cal E} $ vs.\@ light cone time $ v $ for $(l_1, l_2) = (0.50,0.50), (0.45,0.55), (0.43,0.57)$ and $(0.41,0.59)$, which are shown by blue, green, red and black curves respectively. Here, $ l_s = 0.50$, $ {\cal E} $ is in units of $ \frac{3}{16 G_N^{(3)}} $.}
 \label{vaidya_ads_plot_en_vs_v}
\end{center}
\end{figure}

In figure \ref{vaidya_ads_plot_en_vs_v}, we plot the entanglement negativity as a function of $v$ for dif{}ferent sizes of the intervals. For this we have chosen the mass function as $ m(v) = \tanh v $ as described in \cite{Hubeny:2007xt,AbajoArrastia:2010yt}. In all the cases, the entanglement negativity decreases monotonically with $ v $ and eventually saturates for large $ v $, which characterizes the thermalization of the mixed state under consideration. This thermalization corresponds to the black hole formation in the bulk. This result is completely consistent with quantum information theory expectations \cite{Zurek:2003zz}.
%
%
%
%
In quantum information theory an entanglement measure is expected to be non-negative and a monotonically decreasing function of the temperature \cite{Zurek:2003zz}. The behavior of the entanglement negativity in figure \ref{vaidya_ads_plot_en_vs_v} is consistent with the above expectations and the fact that it characterizes the upper bound on the distillable entanglement for mixed states.
This is in contrast to the corresponding behavior of the holographic entanglement entropy involving the bulk Vaidya-$AdS_3$ geometry, where it increases monotonically and saturates to the value of the thermal entropy for large $ v $ \cite{Hubeny:2007xt}. At finite temperatures, the entanglement entropy receives contributions from both quantum and thermal correlations, and is dominated by the thermal correlations at high temperatures. Hence it increases monotonically as the mixed state thermalizes and saturates to the value of the thermal entropy for large $ v $.

\section{Summary and conclusions}
\label{sec_summary}

To summarize we have established a covariant holographic entanglement negativity construction for time dependent mixed state configurations of disjoint subsystems in proximity in $(1+1)$ dimensional conformal field theories dual to non static bulk $AdS_3$ geometries. To this end a large central charge analysis for the corresponding replica technique results in the dual $CFT_{1+1}$ has been utilized. Our construction involves a specific algebraic sum of the lengths of bulk extremal curves ( geodesics) anchored on appropriate combinations of the intervals. This may also be expressed in terms of the holographic mutual informations between specific combinations of the intervals involved.

In this context we have obtained the holographic entanglement negativity for the mixed state configurations of disjoint intervals in proximity in $CFT_{1+1}$s dual to specific bulk $AdS_3$ geometries. These include $CFT_{1+1}$s with a conserved charge, dual to stationary bulk $AdS_3$ configurations described by non extremal and extremal rotating BTZ black holes. Our results for these cases match exactly with the corresponding replica technique results in the large central charge limit and constitute strong consistency checks for our construction.
Interestingly the covariant holographic entanglement negativity decouples into two parts corresponding to the left and right moving temperatures in the finite temperature $CFT_{1+1}$ dual to the bulk non extremal rotating BTZ black hole. For the extremal rotating BTZ case on the other hand, it decouples into the entanglement negativity for the ground state in a $CFT_{1+1}$ dual to a bulk pure $AdS_3$ geometry and that for a thermal like state described by a bulk Frolov-Thorne temperature.

Furthermore we have utilized our construction to investigate the time dependent mixed state configuration of disjoint intervals in proximity in a $CFT_{1+1}$ dual to a bulk Vaidya-$AdS_3$ geometry. Remarkably we have been able to demonstrate the time dependent aspect of this configuration where the holographic entanglement negativity monotonically decreases with time leading to its eventual saturation to a small fixed value. This characterizes the thermalization of the mixed state in the dual $CFT_{1+1}$ under consideration, and corresponds to the formation of a bulk black hole.

We emphasize that our covariant holographic entanglement negativity conjecture furnishes an elegant method to compute the entanglement negativity for time dependent mixed state configurations in dual $CFT_{1+1}$s. However although substantiated with non trivial examples, a bulk proof for our construction along the lines of \cite{Faulkner:2013yia} is an outstanding issue. Our holographic construction also suggests a higher dimensional extension to a generic $AdS_{d+1}/CFT_d$ scenario involving co dimension two bulk extremal surfaces anchored on appropriate subsystems. Naturally such an extension would also require substantive examples and a bulk proof along the lines of \cite{Lewkowycz:2013nqa,Dong:2016hjy}.

Clearly our holographic construction proposed here will have significant impact for the investigation of time dependent entanglement issues such as time evolution of entanglement in quantum quenches and thermalization, black hole formation and collapse scenarios, information loss paradox, unitarity restoration and the related firewall problem in the context of $AdS_3/CFT_2$ holography. These constitute interesting open issues for future investigations.


\bibliographystyle{utphys}

\bibliography{cohendj_bib}

\providecommand{\href}[2]{#2}\begingroup\raggedright\begin{thebibliography}{10}

\bibitem{Calabrese:2004eu}
P.~Calabrese and J.~L. Cardy, ``{Entanglement entropy and quantum field
  theory},'' \href{http://dx.doi.org/10.1088/1742-5468/2004/06/P06002}{{\em J.
  Stat. Mech.} {\bfseries 0406} (2004) P06002},
\href{http://arxiv.org/abs/hep-th/0405152}{{\ttfamily arXiv:hep-th/0405152
  [hep-th]}}.

\bibitem{Calabrese:2009qy}
P.~Calabrese and J.~Cardy, ``{Entanglement entropy and conformal field
  theory},'' \href{http://dx.doi.org/10.1088/1751-8113/42/50/504005}{{\em J.
  Phys.} {\bfseries A42} (2009) 504005},
\href{http://arxiv.org/abs/0905.4013}{{\ttfamily arXiv:0905.4013
  [cond-mat.stat-mech]}}.

\bibitem{Cardy:2007mb}
J.~L. Cardy, O.~A. Castro-Alvaredo, and B.~Doyon, ``{Form factors of
  branch-point twist fields in quantum integrable models and entanglement
  entropy},'' \href{http://dx.doi.org/10.1007/s10955-007-9422-x}{{\em J.
  Statist. Phys.} {\bfseries 130} (2008) 129--168},
\href{http://arxiv.org/abs/0706.3384}{{\ttfamily arXiv:0706.3384 [hep-th]}}.

\bibitem{1987NuPhB.282...13D}
L.~{Dixon}, D.~{Friedan}, E.~{Martinec}, and S.~{Shenker}, ``{The conformal
  field theory of orbifolds},''
  \href{http://dx.doi.org/10.1016/0550-3213(87)90676-6}{{\em Nuclear Physics B}
  {\bfseries 282} (1987) 13--73}.

\bibitem{1987CMaPh.112..567K}
V.~G. {Knizhnik}, ``{Analytic fields on Riemann surfaces. II},''
  \href{http://dx.doi.org/10.1007/BF01225373}{{\em Communications in
  Mathematical Physics} {\bfseries 112} (Dec., 1987) 567--590}.

\bibitem{Vidal:2002zz}
G.~Vidal and R.~F. Werner, ``{Computable measure of entanglement},''
  \href{http://dx.doi.org/10.1103/PhysRevA.65.032314}{{\em Phys. Rev.}
  {\bfseries A65} (2002) 032314},
\href{http://arxiv.org/abs/quant-ph/0102117}{{\ttfamily arXiv:quant-ph/0102117
  [quant-ph]}}.

\bibitem{Plenio:2005cwa}
M.~B. Plenio, ``{Logarithmic Negativity: A Full Entanglement Monotone That is
  not Convex},'' \href{http://dx.doi.org/10.1103/PhysRevLett.95.090503}{{\em
  Phys. Rev. Lett.} {\bfseries 95} no.~9, (2005) 090503},
\href{http://arxiv.org/abs/quant-ph/0505071}{{\ttfamily arXiv:quant-ph/0505071
  [quant-ph]}}.

\bibitem{Calabrese:2012ew}
P.~Calabrese, J.~Cardy, and E.~Tonni, ``{Entanglement negativity in quantum
  field theory},'' \href{http://dx.doi.org/10.1103/PhysRevLett.109.130502}{{\em
  Phys. Rev. Lett.} {\bfseries 109} (2012) 130502},
\href{http://arxiv.org/abs/1206.3092}{{\ttfamily arXiv:1206.3092
  [cond-mat.stat-mech]}}.

\bibitem{Calabrese:2012nk}
P.~Calabrese, J.~Cardy, and E.~Tonni, ``{Entanglement negativity in extended
  systems: A field theoretical approach},''
  \href{http://dx.doi.org/10.1088/1742-5468/2013/02/P02008}{{\em J. Stat.
  Mech.} {\bfseries 1302} (2013) P02008},
\href{http://arxiv.org/abs/1210.5359}{{\ttfamily arXiv:1210.5359
  [cond-mat.stat-mech]}}.

\bibitem{Calabrese:2014yza}
P.~Calabrese, J.~Cardy, and E.~Tonni, ``{Finite temperature entanglement
  negativity in conformal field theory},''
  \href{http://dx.doi.org/10.1088/1751-8113/48/1/015006}{{\em J. Phys.}
  {\bfseries A48} no.~1, (2015) 015006},
\href{http://arxiv.org/abs/1408.3043}{{\ttfamily arXiv:1408.3043
  [cond-mat.stat-mech]}}.

\bibitem{Coser:2014gsa}
A.~Coser, E.~Tonni, and P.~Calabrese, ``{Entanglement negativity after a global
  quantum quench},''
  \href{http://dx.doi.org/10.1088/1742-5468/2014/12/P12017}{{\em J. Stat.
  Mech.} {\bfseries 1412} no.~12, (2014) P12017},
\href{http://arxiv.org/abs/1410.0900}{{\ttfamily arXiv:1410.0900
  [cond-mat.stat-mech]}}.

\bibitem{Hoogeveen:2014bqa}
M.~Hoogeveen and B.~Doyon, ``{Entanglement negativity and entropy in
  non-equilibrium conformal field theory},''
  \href{http://dx.doi.org/10.1016/j.nuclphysb.2015.06.021}{{\em Nucl. Phys.}
  {\bfseries B898} (2015) 78--112},
\href{http://arxiv.org/abs/1412.7568}{{\ttfamily arXiv:1412.7568
  [cond-mat.stat-mech]}}.

\bibitem{2014NJPh...16l3020E}
V.~{Eisler} and Z.~{Zimbor{\'a}s}, ``{Entanglement negativity in the harmonic
  chain out of equilibrium},''
  \href{http://dx.doi.org/10.1088/1367-2630/16/12/123020}{{\em New Journal of
  Physics} {\bfseries 16} no.~12, (Dec., 2014) 123020},
  \href{http://arxiv.org/abs/1406.5474}{{\ttfamily arXiv:1406.5474
  [cond-mat.stat-mech]}}.

\bibitem{Wen:2015qwa}
X.~Wen, P.-Y. Chang, and S.~Ryu, ``{Entanglement negativity after a local
  quantum quench in conformal field theories},''
  \href{http://dx.doi.org/10.1103/PhysRevB.92.075109}{{\em Phys. Rev.}
  {\bfseries B92} no.~7, (2015) 075109},
\href{http://arxiv.org/abs/1501.00568}{{\ttfamily arXiv:1501.00568
  [cond-mat.stat-mech]}}.

\bibitem{Ryu:2006bv}
S.~Ryu and T.~Takayanagi, ``{Holographic derivation of entanglement entropy
  from AdS/CFT},'' \href{http://dx.doi.org/10.1103/PhysRevLett.96.181602}{{\em
  Phys. Rev. Lett.} {\bfseries 96} (2006) 181602},
\href{http://arxiv.org/abs/hep-th/0603001}{{\ttfamily arXiv:hep-th/0603001
  [hep-th]}}.

\bibitem{Ryu:2006ef}
S.~Ryu and T.~Takayanagi, ``{Aspects of Holographic Entanglement Entropy},''
  \href{http://dx.doi.org/10.1088/1126-6708/2006/08/045}{{\em JHEP} {\bfseries
  08} (2006) 045},
\href{http://arxiv.org/abs/hep-th/0605073}{{\ttfamily arXiv:hep-th/0605073
  [hep-th]}}.

\bibitem{Nishioka:2009un}
T.~Nishioka, S.~Ryu, and T.~Takayanagi, ``{Holographic Entanglement Entropy: An
  Overview},'' \href{http://dx.doi.org/10.1088/1751-8113/42/50/504008}{{\em J.
  Phys.} {\bfseries A42} (2009) 504008},
\href{http://arxiv.org/abs/0905.0932}{{\ttfamily arXiv:0905.0932 [hep-th]}}.

\bibitem{Takayanagi:2012kg}
T.~Takayanagi, ``{Entanglement Entropy from a Holographic Viewpoint},''
  \href{http://dx.doi.org/10.1088/0264-9381/29/15/153001}{{\em Class. Quant.
  Grav.} {\bfseries 29} (2012) 153001},
\href{http://arxiv.org/abs/1204.2450}{{\ttfamily arXiv:1204.2450 [gr-qc]}}.

\bibitem{Nishioka:2018khk}
T.~Nishioka, ``{Entanglement entropy: holography and renormalization group},''
\href{http://arxiv.org/abs/1801.10352}{{\ttfamily arXiv:1801.10352 [hep-th]}}.

\bibitem{Hubeny:2007xt}
V.~E. Hubeny, M.~Rangamani, and T.~Takayanagi, ``{A Covariant holographic
  entanglement entropy proposal},''
  \href{http://dx.doi.org/10.1088/1126-6708/2007/07/062}{{\em JHEP} {\bfseries
  07} (2007) 062},
\href{http://arxiv.org/abs/0705.0016}{{\ttfamily arXiv:0705.0016 [hep-th]}}.

\bibitem{Bousso:1999xy}
R.~Bousso, ``{A Covariant entropy conjecture},''
  \href{http://dx.doi.org/10.1088/1126-6708/1999/07/004}{{\em JHEP} {\bfseries
  07} (1999) 004},
\href{http://arxiv.org/abs/hep-th/9905177}{{\ttfamily arXiv:hep-th/9905177
  [hep-th]}}.

\bibitem{Bousso:1999cb}
R.~Bousso, ``{Holography in general space-times},''
  \href{http://dx.doi.org/10.1088/1126-6708/1999/06/028}{{\em JHEP} {\bfseries
  06} (1999) 028},
\href{http://arxiv.org/abs/hep-th/9906022}{{\ttfamily arXiv:hep-th/9906022
  [hep-th]}}.

\bibitem{Bousso:2002ju}
R.~Bousso, ``{The Holographic principle},''
  \href{http://dx.doi.org/10.1103/RevModPhys.74.825}{{\em Rev. Mod. Phys.}
  {\bfseries 74} (2002) 825--874},
\href{http://arxiv.org/abs/hep-th/0203101}{{\ttfamily arXiv:hep-th/0203101
  [hep-th]}}.

\bibitem{Balasubramanian:2011ur}
V.~Balasubramanian, A.~Bernamonti, J.~de~Boer, N.~Copland, B.~Craps,
  E.~Keski-Vakkuri, B.~Muller, A.~Schafer, M.~Shigemori, and W.~Staessens,
  ``{Holographic Thermalization},''
  \href{http://dx.doi.org/10.1103/PhysRevD.84.026010}{{\em Phys. Rev.}
  {\bfseries D84} (2011) 026010},
\href{http://arxiv.org/abs/1103.2683}{{\ttfamily arXiv:1103.2683 [hep-th]}}.

\bibitem{Hartman:2013qma}
T.~Hartman and J.~Maldacena, ``{Time Evolution of Entanglement Entropy from
  Black Hole Interiors},''
  \href{http://dx.doi.org/10.1007/JHEP05(2013)014}{{\em JHEP} {\bfseries 05}
  (2013) 014},
\href{http://arxiv.org/abs/1303.1080}{{\ttfamily arXiv:1303.1080 [hep-th]}}.

\bibitem{Albash:2010mv}
T.~Albash and C.~V. Johnson, ``{Evolution of Holographic Entanglement Entropy
  after Thermal and Electromagnetic Quenches},''
  \href{http://dx.doi.org/10.1088/1367-2630/13/4/045017}{{\em New J. Phys.}
  {\bfseries 13} (2011) 045017},
\href{http://arxiv.org/abs/1008.3027}{{\ttfamily arXiv:1008.3027 [hep-th]}}.

\bibitem{Caputa:2013eka}
P.~Caputa, G.~Mandal, and R.~Sinha, ``{Dynamical entanglement entropy with
  angular momentum and U(1) charge},''
  \href{http://dx.doi.org/10.1007/JHEP11(2013)052}{{\em JHEP} {\bfseries 11}
  (2013) 052},
\href{http://arxiv.org/abs/1306.4974}{{\ttfamily arXiv:1306.4974 [hep-th]}}.

\bibitem{Mandal:2016cdw}
G.~Mandal, R.~Sinha, and T.~Ugajin, ``{Finite size effect on dynamical
  entanglement entropy: CFT and holography},''
\href{http://arxiv.org/abs/1604.07830}{{\ttfamily arXiv:1604.07830 [hep-th]}}.

\bibitem{David:2016pzn}
J.~R. David, S.~Khetrapal, and S.~P. Kumar, ``{Universal corrections to
  entanglement entropy of local quantum quenches},''
  \href{http://dx.doi.org/10.1007/JHEP08(2016)127}{{\em JHEP} {\bfseries 08}
  (2016) 127},
\href{http://arxiv.org/abs/1605.05987}{{\ttfamily arXiv:1605.05987 [hep-th]}}.

\bibitem{Chaturvedi:2016rcn}
P.~Chaturvedi, V.~Malvimat, and G.~Sengupta, ``{Holographic Quantum
  Entanglement Negativity},''
  \href{http://dx.doi.org/10.1007/JHEP05(2018)172}{{\em JHEP} {\bfseries 05}
  (2018) 172},
\href{http://arxiv.org/abs/1609.06609}{{\ttfamily arXiv:1609.06609 [hep-th]}}.

\bibitem{Chaturvedi:2016rft}
P.~Chaturvedi, V.~Malvimat, and G.~Sengupta, ``{Entanglement negativity,
  Holography and Black holes},''
  \href{http://dx.doi.org/10.1140/epjc/s10052-018-5969-8}{{\em Eur. Phys. J.}
  {\bfseries C78} no.~6, (2018) 499},
\href{http://arxiv.org/abs/1602.01147}{{\ttfamily arXiv:1602.01147 [hep-th]}}.

\bibitem{Chaturvedi:2016opa}
P.~Chaturvedi, V.~Malvimat, and G.~Sengupta, ``{Covariant holographic
  entanglement negativity},''
  \href{http://dx.doi.org/10.1140/epjc/s10052-018-6259-1}{{\em Eur. Phys. J.}
  {\bfseries C78} no.~9, (2018) 776},
\href{http://arxiv.org/abs/1611.00593}{{\ttfamily arXiv:1611.00593 [hep-th]}}.

\bibitem{Malvimat:2017yaj}
V.~Malvimat and G.~Sengupta, ``{Entanglement negativity at large central
  charge},''
\href{http://arxiv.org/abs/1712.02288}{{\ttfamily arXiv:1712.02288 [hep-th]}}.

\bibitem{Headrick:2010zt}
M.~Headrick, ``{Entanglement Renyi entropies in holographic theories},''
  \href{http://dx.doi.org/10.1103/PhysRevD.82.126010}{{\em Phys. Rev.}
  {\bfseries D82} (2010) 126010},
\href{http://arxiv.org/abs/1006.0047}{{\ttfamily arXiv:1006.0047 [hep-th]}}.

\bibitem{Hartman:2013mia}
T.~Hartman, ``{Entanglement Entropy at Large Central Charge},''
\href{http://arxiv.org/abs/1303.6955}{{\ttfamily arXiv:1303.6955 [hep-th]}}.

\bibitem{Fitzpatrick:2014vua}
A.~L. Fitzpatrick, J.~Kaplan, and M.~T. Walters, ``{Universality of
  Long-Distance AdS Physics from the CFT Bootstrap},''
  \href{http://dx.doi.org/10.1007/JHEP08(2014)145}{{\em JHEP} {\bfseries 08}
  (2014) 145},
\href{http://arxiv.org/abs/1403.6829}{{\ttfamily arXiv:1403.6829 [hep-th]}}.

\bibitem{Faulkner:2013yia}
T.~Faulkner, ``{The Entanglement Renyi Entropies of Disjoint Intervals in
  AdS/CFT},''
\href{http://arxiv.org/abs/1303.7221}{{\ttfamily arXiv:1303.7221 [hep-th]}}.

\bibitem{Lewkowycz:2013nqa}
A.~Lewkowycz and J.~Maldacena, ``{Generalized gravitational entropy},''
  \href{http://dx.doi.org/10.1007/JHEP08(2013)090}{{\em JHEP} {\bfseries 08}
  (2013) 090},
\href{http://arxiv.org/abs/1304.4926}{{\ttfamily arXiv:1304.4926 [hep-th]}}.

\bibitem{Jain:2017sct}
P.~Jain, V.~Malvimat, S.~Mondal, and G.~Sengupta, ``{Holographic entanglement
  negativity conjecture for adjacent intervals in $AdS_3/CFT_2$},''
  \href{http://dx.doi.org/10.1016/j.physletb.2019.04.037}{{\em Phys. Lett.}
  {\bfseries B793} (2019) 104--109},
\href{http://arxiv.org/abs/1707.08293}{{\ttfamily arXiv:1707.08293 [hep-th]}}.

\bibitem{Jain:2017xsu}
P.~Jain, V.~Malvimat, S.~Mondal, and G.~Sengupta, ``{Holographic entanglement
  negativity for adjacent subsystems in AdS$_{d+1}$/CFT$_{d}$},''
  \href{http://dx.doi.org/10.1140/epjp/i2018-12113-0}{{\em Eur. Phys. J. Plus}
  {\bfseries 133} no.~8, (2018) 300},
\href{http://arxiv.org/abs/1708.00612}{{\ttfamily arXiv:1708.00612 [hep-th]}}.

\bibitem{Jain:2018bai}
P.~Jain, V.~Malvimat, S.~Mondal, and G.~Sengupta, ``{Holographic Entanglement
  Negativity for Conformal Field Theories with a Conserved Charge},''
  \href{http://dx.doi.org/10.1140/epjc/s10052-018-6383-y}{{\em Eur. Phys. J.}
  {\bfseries C78} no.~11, (2018) 908},
\href{http://arxiv.org/abs/1804.09078}{{\ttfamily arXiv:1804.09078 [hep-th]}}.

\bibitem{Jain:2017uhe}
P.~Jain, V.~Malvimat, S.~Mondal, and G.~Sengupta, ``{Covariant Holographic
  Entanglement Negativity Conjecture for Adjacent Subsystems in
  $\mathrm{AdS_{3}/CFT_2}$},''
\href{http://arxiv.org/abs/1710.06138}{{\ttfamily arXiv:1710.06138 [hep-th]}}.

\bibitem{Malvimat:2018txq}
V.~Malvimat, S.~Mondal, B.~Paul, and G.~Sengupta, ``{Holographic entanglement
  negativity for disjoint intervals in $AdS_3/CFT_2$},''
  \href{http://dx.doi.org/10.1140/epjc/s10052-019-6693-8}{{\em Eur. Phys. J.}
  {\bfseries C79} no.~3, (2019) 191},
\href{http://arxiv.org/abs/1810.08015}{{\ttfamily arXiv:1810.08015 [hep-th]}}.

\bibitem{Anous:2017tza}
T.~Anous, T.~Hartman, A.~Rovai, and J.~Sonner, ``{From Conformal Blocks to Path
  Integrals in the Vaidya Geometry},''
  \href{http://dx.doi.org/10.1007/JHEP09(2017)009}{{\em JHEP} {\bfseries 09}
  (2017) 009},
\href{http://arxiv.org/abs/1706.02668}{{\ttfamily arXiv:1706.02668 [hep-th]}}.

\bibitem{Kulaxizi:2014nma}
M.~Kulaxizi, A.~Parnachev, and G.~Policastro, ``{Conformal Blocks and
  Negativity at Large Central Charge},''
  \href{http://dx.doi.org/10.1007/JHEP09(2014)010}{{\em JHEP} {\bfseries 09}
  (2014) 010},
\href{http://arxiv.org/abs/1407.0324}{{\ttfamily arXiv:1407.0324 [hep-th]}}.

\bibitem{Brown:1986nw}
J.~D. Brown and M.~Henneaux, ``{Central Charges in the Canonical Realization of
  Asymptotic Symmetries: An Example from Three-Dimensional Gravity},''
\href{http://dx.doi.org/10.1007/BF01211590}{{\em Commun. Math. Phys.}
  {\bfseries 104} (1986) 207--226}.

\bibitem{Frolov:1989jh}
V.~P. Frolov and K.~S. Thorne, ``{Renormalized Stress - Energy Tensor Near the
  Horizon of a Slowly Evolving, Rotating Black Hole},''
\href{http://dx.doi.org/10.1103/PhysRevD.39.2125}{{\em Phys. Rev.} {\bfseries
  D39} (1989) 2125--2154}.

\bibitem{Caputa:2013lfa}
P.~Caputa, V.~Jejjala, and H.~Soltanpanahi, ``{Entanglement entropy of extremal
  BTZ black holes},'' \href{http://dx.doi.org/10.1103/PhysRevD.89.046006}{{\em
  Phys. Rev.} {\bfseries D89} no.~4, (2014) 046006},
\href{http://arxiv.org/abs/1309.7852}{{\ttfamily arXiv:1309.7852 [hep-th]}}.

\bibitem{AbajoArrastia:2010yt}
J.~Abajo-Arrastia, J.~Aparicio, and E.~Lopez, ``{Holographic Evolution of
  Entanglement Entropy},''
  \href{http://dx.doi.org/10.1007/JHEP11(2010)149}{{\em JHEP} {\bfseries 11}
  (2010) 149},
\href{http://arxiv.org/abs/1006.4090}{{\ttfamily arXiv:1006.4090 [hep-th]}}.

\bibitem{Zurek:2003zz}
W.~H. Zurek, ``{Decoherence, einselection, and the quantum origins of the
  classical},'' \href{http://dx.doi.org/10.1103/RevModPhys.75.715}{{\em Rev.
  Mod. Phys.} {\bfseries 75} (2003) 715--775},
\href{http://arxiv.org/abs/quant-ph/0105127}{{\ttfamily arXiv:quant-ph/0105127
  [quant-ph]}}.

\bibitem{Dong:2016hjy}
X.~Dong, A.~Lewkowycz, and M.~Rangamani, ``{Deriving covariant holographic
  entanglement},'' \href{http://dx.doi.org/10.1007/JHEP11(2016)028}{{\em JHEP}
  {\bfseries 11} (2016) 028},
\href{http://arxiv.org/abs/1607.07506}{{\ttfamily arXiv:1607.07506 [hep-th]}}.

\end{thebibliography}\endgroup

\end{document}